\documentclass[twocolumn,prl,showpacs]{revtex4}
\usepackage{graphicx}
\usepackage{dcolumn}
\usepackage{amssymb}

\def\dd{\ensuremath{\mathrm{d}}}
\def\MZ{\ensuremath{\mathrm{MZ}}}
\def\Ek{E_{\mathrm{k}}}
\def\mP{m_{\mathrm{P}}}
\def\lP{\ell_{\mathrm{P}}}
\def\gw{\mathrm{gw}}

\begin{document}

\title{Ultimate decoherence border for matter-wave interferometry}
\author{Brahim Lamine}
\email{lamine@spectro.jussieu.fr}
\author{R\'emy Herv\'e}
\author{Astrid Lambrecht}
\author{Serge Reynaud}
\affiliation{Laboratoire Kastler Brossel,
\footnote{Unit\'e mixte du CNRS, de l'ENS et de l'UPMC.}
Universit\'e Pierre et Marie Curie, case74,
Campus Jussieu, F-75252 Paris cedex 05}

\date{\today}

\begin{abstract}
Stochastic backgrounds of gravitational waves are intrinsic fluctuations of spacetime which lead to an unavoidable decoherence mechanism.
This mechanism manifests itself as a degradation of the contrast of quantum interferences. It defines an ultimate decoherence border for
matter-wave interferometry using larger and larger molecules. We give a quantitative characterization of this border in terms of figures
involving the gravitational environment as well as the sensitivity of the interferometer to gravitational waves. The known level of
gravitational noise determines the maximal size of the molecular probe for which interferences may remain observable. We discuss the
relevance of this result in the context of ongoing progresses towards more and more sensitive matter-wave interferometry.
\end{abstract}


\maketitle

Fluctuations of spacetime are often referred to as a natural source of decoherence that defines an ultimate border for quantum
interferences. The idea, evoked long ago by Feynman~\cite{feynman}, relies on the fact that the Planck mass $\mP =\sqrt{\hbar c/G}$ built on
the Planck constant $\hbar$, the velocity of light $c$ and the Newton constant $G$, has a value $\simeq 22\,\mu$g which lies on the
borderland between microscopic and macroscopic masses. An object with a mass $m$ larger than $\mP$ is thus associated to a Compton
wavelength $\hbar /mc$ smaller than the Planck length $\lP=\hbar /\mP c$ typical of quantum fuzziness of spacetime. Though this length
scale $\lP\sim10^{-35}\,$m is not directly accessible to experiments, one may wonder whether fluctuation behaviours are modified when $m$
crosses the mass scale $\mP$~\cite{karolyhazy:66,penrose:96,jaekel:94}.

In this letter, we make this qualitative argument more specific by considering matter-wave interferometers as the quantum system
and stochastic backgrounds of gravitational waves (GW) as the source of their decoherence. Decoherence mechanism which might arise
from Planck scale fluctuations of spacetime have already been studied in the literature~\cite{percival,amelino}, with however a
large uncertainty on the level of the latter fluctuations. Here, we focus our attention on known sources of spacetime fluctuations,
namely GW backgrounds predicted by general relativity to be generated by astrophysical or cosmological processes.
This source of noise, to be discussed in more detail later on, leads to an intrinsic decoherence mechanism against which
interferometers cannot be shielded~\cite{reynaud:2001:2002,lamine:2002,lamine:thesis}.
We give a quantitative characterization of this mechanism in terms of relevant figures built up on the spectrum
of the gravitational noise and the sensitivity of matter-wave interferometers to this noise.

Our main purpose is to apply these ideas to the more and more sensitive matter-wave interferometers presently developed with larger and
larger molecules~\cite{zeilinger}. At the moment, the decoherence of these instruments stems from collisions with the residual gas, emission
of thermal radiation by the molecules or instrumental dephasings produced for example by vibrations of the mechanical structure. These noise
sources can in principle be reduced by using higher vacuum, lower temperature, improved velocity selection and, more generally, a better
controlled and quieter environment available in particular in space experiments~\cite{Hyper00}. With the ongoing rapid progress in this
domain, more fundamental limits may eventually be reached. It is precisely the border induced by gravitationally induced decoherence which
is investigated in the present letter. In particular, we compute the maximal mass for a molecular probe that preserves interferences. In the
quantitative study presented here, this mass does not only depend on the Planck mass, but also on the geometry of the interferometer and on
the gravitational noise level. Hence, this result brings the qualitative argument of Feynman to a quantitative estimation for a specific
well defined physical problem.

Besides those GW bursts which are looked for by interferometric detectors~\cite{gwdetectors}, there exist stochastic GW backgrounds
extending over a large frequency range. A first part of this background is originating from the gravitational emission of binary systems in
our galaxy and its vicinity~\cite{bender:97}. The stochastic character thus comes from our lack of knowledge on the precise parameters
associated with the enormous number of unresolved binaries. A second part of the background has a cosmological origin coming from the
primordial era of the cosmic evolution. These relic GW are produced by an amplification, occuring during the expansion of the Universe, of
the primordial vacuum fluctuations of the gravitational field~\cite{grishchuk}. For the sake of simplicity, we will suppose these
backgrounds to be gaussian, stationary, unpolarized and isotropic. These simplifying assumptions are indeed sufficient for giving an
estimation of decoherence induced by the scattering of these two backgrounds.

Within this context, the backgrounds are described by a spectral density $S_h[\omega]$ of strain fluctuations at a given frequency $\omega$.
This function is equivalent to the correlation of the metric in a Transverse Traceless (TT) coordinate system, at a fixed position taken
here to be the center of the TT coordinate system~:
\begin{eqnarray}
\label{CorrelationFunctions}
&&\left<h_{ij}(t)h_{kl}(0)\right>= \delta_{ijkl} \int\frac{\dd\omega}{2\pi}\,S_h[\omega]e^{-i\omega t} \\
&&\delta_{ijkl} \equiv \delta_{ik}\delta_{jl}+\delta_{il}\delta_{jk}-\frac{2}{3}\delta_{ij}\delta_{kl} \nonumber
\end{eqnarray}
Informations on the spectral density $S_h[\omega]$ can be found in~\cite{schutz:99,maggiore:00,ungarelli:01,grishchuk:01}. The binary
confusion background, deduced from the known distribution of binaries in the galaxy, shows a roughly flat plateau between the
$\mu$Hz and mHz and drops rapidly on both sides on this plateau. The cosmological background shows a $1/\omega^3$ dependence and should
dominate at low frequencies. It depends on a poorly known parameter $\Omega_\gw$ measuring the GW energy density compared to the critical
cosmic density.

Those GW induce a distortion of the interferometric paths and therefore lead to a differential phase shift which can be evaluated simply in
the eikonal approximation~\cite{tourrenc:76}. In this approximation the Hamilton-Jacobi theory leads to an identification of the matter-wave
phase $\Phi$ to the action $S$ divided by $\hbar$. The action corresponds to a Lagrangian density~\cite{landau:book} which couples the
spatial part of the metric $h_{ij}$ to the second derivative of the quadrupole $Q^{ij}$ of the interferometer~\cite{weinberg:book}~:
\begin{eqnarray}
\label{ActionQuadrupolaire}
&&S=\frac{1}{4}\int\dd t\;h_{ij}(t) {\dd^2{Q}^{ij}(t) \over \dd t^2}\\
&&Q^{ij}(t)=\frac{1}{c^2}\int\dd^3\mathbf{x}\left(x^ix^j-\frac{1}{3}
\,\delta^{ij}x^kx_k\right) T^{00}(t,\mathbf{x}) \nonumber
\end{eqnarray}

This quadrupole coupling is equivalent to the dipole approximation used in electromagnetism to describe the coupling on an atom having a
size much smaller than the wavelength. The dephasing $\Delta\Phi$ between the two arms of the interferometer is given by the difference
$\Delta S$ between the two action integrals and, then, by the expression (\ref{ActionQuadrupolaire}) with $Q^{ij}$ replaced by the
difference $\Delta Q^{ij}$ between the quadrupoles evaluated when the probe follows either one arm or the other one.
The resulting expression may be written in the frequency domain as~:
\begin{eqnarray}
\label{Dephasing}
&&\Delta\Phi(t)=\frac{\Delta S(t)}{\hbar}\equiv\int\frac{\dd\omega}{2\pi}h_{ij}[\omega]a^{ij}[\omega]e^{-i\omega t} \\
&&a^{ij}[\omega]={i\over 4\hbar} \int\frac{\dd\omega^\prime}{2\pi}\,\omega^{\prime\,2}\Delta Q^{ij}[\omega^\prime]\,
\frac{1-e^{-i(\omega+\omega^\prime)\tau}}{\omega+\omega^\prime}\nonumber
\end{eqnarray}
The apparatus function $a^{ij}[\omega]$ describes the geometry of the interferometer which has been assumed here to have a rhombic form with
$\tau$ the time of flight along each arm. Note that the previous expression of $a^{ij}$ is valid for atomic interferometers where mirrors
and beam splitters for atoms are built up on laser beams and perceived by atoms as freely falling objects~\cite{lamine:thesis}.

The expression (\ref{Dephasing}) of the gravitational phase shift is an integral over the whole frequency spectrum. It contains two effects
corresponding respectively to a global phase shift of the interferogram and to a reduction of its contrast. The separation of these two
effects relies on the comparison of the GW frequency with the inverse of the measuring time $T$. The precise definition of this time $T$
requires a complete study of a specific model of interferometer. In the general discussion presented here, we will define it as the minimal
time needed to build an interferogram. $T$ is not only larger than the time of flight $\tau$ of the atoms along the two arms of the
interferometer, but it has often to be much larger than $\tau$ in order to reach a signal to noise ratio sufficient to see the interferogram
figure. While the signal to noise ratio is improved by averaging over a longer time $T$, the contrast of the interferogram is decreased by
the change of the gravitational environment. The decoherence mechanism studied in this letter is precisely the result of this potential
blurring of the fringes which would occur before the fringes have even became visible, should the noise be too large.

Formal equivalence between the loss of contrast of interference fringes and the general theory of
decoherence has been established in~\cite{imry:90,lamine:thesis}. It turns out that the reduction of the contrast
exactly corresponds to the trace over the unobserved degrees of freedom of the gravitational environment.
``Unobserved'' here refers to gravitational noise lying outside the frequency window where fluctuations can be detected.
In the following, we will denote $\delta\varphi$ the contribution of this uncontrolled noise which leads to a degradation
of the contrast for a given signal processing strategy. We will also consider that this $\delta\varphi$ is deduced from
$\Delta\Phi$ through a filtering function $f$ in the frequency domain~:
\begin{equation}
\label{Filter}
\delta\varphi(t)=\int\frac{\dd\omega}{2\pi}h_{ij}[\omega]a^{ij}[\omega]f[\omega]e^{-i\omega t}
\end{equation}
In the simple signal processing strategy which consists in averaging the interferogram over a measuring time $T$,
the function $f[\omega]$ is just the high pass filter defining uncontrolled noise as frequencies larger than the inverse of $T$.
More elaborated signal processing strategies could be studied by defining more general functions $f$.
Decoherence is then characterized by the value of the fringe contrast $\mathcal{V}$ deduced, within a gaussian description of the
fluctuations, as the exponential of the variance of the uncontrolled noise~\cite{lamine:thesis}~:
\begin{equation}
\mathcal{V}=\left< \mathrm{exp}(i\delta\varphi)\right>=\mathrm{exp} \left(-\frac{\Delta\varphi^2}{2}\right) \, ,\quad
\Delta\varphi^2=\left<\delta\varphi^2\right> \label{Visibility}
\end{equation}

Using expressions (\ref{Dephasing}-\ref{Filter}) of the phase noises as well as the correlation functions (\ref{CorrelationFunctions}) of
the metric perturbation, we finally rewrite the variance $\Delta\varphi^2$ as an integral in the frequency domain~\cite{lamine:2002}~:
\begin{eqnarray}
\label{deltaphi2}
&&\Delta\varphi^2=\int\frac{\dd\omega}{2\pi}S_h[\omega]\mathcal{A}[\omega]F[\omega] \\
F[\omega]&=&|f[\omega]|^2\quad,\quad \mathcal{A}[\omega]=\delta_{ijkl}\,a^{ij}[\omega]a^{kl}[-\omega] \nonumber
\end{eqnarray}
The integrand is the product of three terms, the gravitational noise spectrum $S_h$, the apparatus response function $\mathcal{A}$ and the
filter function $F$.

As already emphasized, $\mathcal{A}[\omega]$ has a complicated expression depending on the geometry of the interferometer. For forthcoming
discussions, we will consider the commonly studied case~\cite{Hyper00} of a Mach-Zehnder geometry with rhombic symmetry in the limits of
small aperture angle ($\alpha\ll1$) and non relativistic velocity ($v\ll c$). The apparatus function is given by the following expression
which captures the main ingredients of the physical description of decoherence~\cite{lamine:2002}~:
\begin{equation}
\mathcal{A}_\MZ[\omega]=\left(4\,\Omega\sin\alpha\right)^2 \left(\frac{1-\cos(\omega\tau)}{\omega}\right)^2\quad
,\quad\Omega=\frac{mv^2}{2\hbar}
\end{equation}
This response function scales as the square of the kinetic energy $\Omega$ of the probe field measured as a frequency. We also remark that
the function $\mathcal{A}_\MZ[\omega]$ goes to zero with the angular separation since the two arms are thus exposed to the same
perturbation. Through its last term finally, it selects a frequency band in the gravitational spectrum which is essentially determined by
the inverse of the time of flight $2\tau$ of the probe field inside the interferometer.

These features are sufficient to give an estimate of decoherence deduced from the preceding calculations~:
\begin{eqnarray}
&&\Delta\varphi^2 \sim \left(4\Omega\tau\sin\alpha\right)^2 \overline{\Delta h^2} \label{FormuleVariance}\\
&&\overline{\Delta h^2} \equiv \int \frac{\dd\omega}{2\pi} S_h[\omega] F[\omega]\left(\frac{1-\cos(\omega\tau)}{\omega\tau}\right)^2
\nonumber
\end{eqnarray}
Assuming that the measuring time $T$ is much larger than $\tau$ and that the filter $F$ cuts off the potential divergence of the integral at
its low frequency side, $\overline{\Delta h^2}$ may essentially be interpreted as the average of the variance $\Delta h^2$ over the
bandwidth $F[\omega]\mathcal{A}[\omega]$.

For a first estimation, we can consider the simple assumption of a bandwidth lying within the plateau of the binary confusion background.
This entails that $S_h$ is roughly flat so that the averaged value $\overline{\Delta h^2}$ is simply given by the noise level on the
plateau, that is $S_h \simeq 10^{-34}\,$s, divided by the time of flight $\tau$. In particular, this leads to a variance
(\ref{FormuleVariance}) that reproduces a Brownian-like diffusion of the phase characterized by a linear dependance of $\Delta\varphi^2$ in
the time of exposition $\tau$ to the perturbation. In the more general discussion that follows, $\overline{\Delta h^2}$ will be computed
with the real spectrum of the GW backgrounds and the real bandwidth of the interferometer.

We now discuss the numbers coming out of expression (\ref{FormuleVariance}) for specific experimental configurations. Our main purpose is to
investigate the possibility for a matter wave interferometer to approach the quantum/classical transition now characterized by the
quantitative condition $\Delta\varphi^2 \sim 1$. Starting from the known fact that $\Delta\varphi^2$ is usually much smaller than unity for
microscopic probes~\cite{lamine:2002}, we see on formula (\ref{FormuleVariance}) that approaching $\Delta\varphi^2 \sim 1$ requires two
kinds of condition. Considering first the point of view of geometry, it is clearly needed to have an interferometer combining large angular
separation and large time of flight. This condition is different from the large area condition helping the atomic interferometer to be used
as an inertial sensor. The difference is due to the fact that GW vary in space and time so that even a null area interferometer could be
sensitive to them. Moreover, sensitivity to GW is determined by the kinetic energy of the probe and not by its rest mass energy. This
entails that rapid probes should be preferred to slow ones, a condition clearly different from the one looked for with atomic
interferometers used as inertial sensors \cite{Hyper00}.

The geometrical and energetical conditions are conflicting with each other~: a high velocity of the matter beams decreases the time of
flight for a given spatial size; meanwhile, it decreases the angular separation between interfering paths if the momentum transfer is fixed,
which is the case for beam splitters built up on Raman scattering processes~\cite{borde:01,chu:01}. This last argument can be made more
precise by introducing $\Delta\Omega=\Delta\Ek/\hbar$ where $\Ek$ is the kinetic energy and $\Delta\Ek$ its variation on the beam splitter,
if we consider the case of a transfered momentum orthogonal to the velocity. With this notation, the variance $\Delta\varphi^2$ is read as
$(4\tau\Delta\Omega)^2 \overline{\Delta h^2}$ which shows that the most relevant parameter for characterizing the beam splitting is
$\Delta\Ek$. When taking as an example the design of the HYPER project~\cite{reynaud:2004}, $\Delta \Ek$ has a very small value of the order
of $10^{-9}\,$eV. This value can be increased by using multiple Raman scattering, up to 140 emissions/absorptions~\cite{chu:94}, but this is
not enough for approaching $\Delta\varphi^2\sim 1$.

The value of $\Delta\Ek$ can be enlarged for example by using magnetic interaction guiding~\cite{andersson:2002}. In this case, it is only
limited by the depth of the guiding well and can go up to a few $10^{-7}\,$eV. Even larger values of the potential depth ($1-100$ meV) are
obtained with non resonant dipole interaction~\cite{pfau:93,houde:00}. Deflection from material gratings also allow high kinetic energy
transfer. For a slit of width $a$ and a diffraction order $n$, the transfered momentum $\hbar\Delta K=\Delta \Ek/v$ has a value of the order
of $2\pi \hbar n/a$ which increases when $a$ decreases. Another promising solution is the inelastic scattering of metastable molecules by
nano-slit transmission gratings~\cite{robert} in which Van Der Waals interactions lead to energy transfer of the order $\Delta\Ek\sim1\,$eV.
For all these configurations however, the energy transfers are still too small to approach the transition $\Delta\varphi^2\sim 1$.

In order to show how challenging is the objective of seeing the quantum/classical transition associated with gravitational decoherence, let
us consider at this point an hypothetical matter-wave interferometer with a wide angle aperture $\sin\alpha\sim1$ and, therefore, a large
kinetic energy transfer $\Delta\Ek\simeq\Ek$. This interferometer can approach the transition $\Delta\varphi^2\sim 1$ if we suppose the beam
to consist of molecules with a mass of $8\times 10^8\,$amu circulating at a velocity $1\,$km s$^{-1}$ in arms with one meter size. These
numbers are calculated with the binary confusion background used as the source of fluctuations. The cosmological contribution would lead to
a slightly weaker effect if we take the value $\Omega_\gw=10^{-14}$. The previous numbers have to be contrasted with advanced projects of
interferometers aiming at large molecules with mass in the $10^5\,$amu scale~\cite{arndt:private}. Meanwhile, they correspond to supersonic
molecular beams with a kinetic energy transfer of the order of $300$~keV, far above the splitting capabilities of the previously mentioned
configurations.

These numbers show that approaching the quantum/classical transition associated with gravitational decoherence is out of reach for the
presently developed molecular interferometers. An attractive idea would be to use Bose Einstein condensates (BEC) instead of large
molecules. Should the BEC respond to the gravitational perturbation as a rigid object containing a large number $N$ of atoms, the parameter
$\Delta\Ek$ would have to be multiplied by the factor $N$ leading to an amplification by a factor $N^2$ of the decoherence rate. A first
characterization of this rigidity condition is that the motion of the center of mass of the BEC induced by gravitational waves should
correspond to frequencies lying well below the excitation spectrum of internal resonances. Detailed calculations are underway to make this
characterization more precise.

Anyway, this argument pleads for BEC used as an interferometric probe in space experiments where the instrumental and environmental noises
can be more efficiently controlled. This could be the best way to test the existence of an ultimate border for the observability of
quantum interferences on matter-wave interferometers, due to the scattering of the spacetime fluctuations.

\begin{acknowledgments}
The authors would like to thank Gert-Ludwig Ingold, Marc-Thierry Jaekel and Paulo Maia-Neto
for many stimulating discussions.
\end{acknowledgments}

\def\and{and }
\def\etal{\textit{et al}}
\newcommand{\Name}[1]{#1,}
\newcommand{\REVIEW}[4]{\textit{#1} \textbf{#2}, #4 (#3)}
\newcommand{\Book}[1]{\textit{#1},}
\newcommand{\Editor}[1]{ed. #1}
\newcommand{\Year}[1]{(#1)}


\begin{thebibliography}{0}


\bibitem{feynman}
  \Name{Feynman R.P. \etal}
  \Book{Feynman lectures on gravitation}
  \Editor{Penguin}
  \Year{1999}

\bibitem{karolyhazy:66}
  \Name{Karolyhazy F.}
  \REVIEW{Nuovo Cim.}{42A}{1966}{390}

\bibitem{penrose:96}
  \Name{Penrose R.}
  \REVIEW{Gen. Rel. Grav.}{28}{1996}{581}

\bibitem{jaekel:94}
  \Name{Jaekel M.-T. \and Reynaud S.}
  \REVIEW{Phys. Lett.}{A185}{1994}{143}

\bibitem{percival}
  \Name{Percival I.C.}
  \REVIEW{Phys. World}{10}{1997}{48};
  \Name{Percival I.C. and Strunz W.T.}
  \REVIEW{Proc. Roy. Soc. London}{A453}{1996}{431}

\bibitem{amelino}
  \Name{Amelino-Camelia G.}
  \REVIEW{Gen. Rel. Grav.}{36}{2004}{539}

\bibitem{reynaud:2001:2002}
  \Name{Reynaud S., Maia-Neto P.A., Lambrecht A. \etal}
  \REVIEW{Europhys. Lett.}{54}{2001}{135};
  \Name{Reynaud S., Lambrecht A., Maia-Neto P. \etal}
  \REVIEW{Int. J. Mod. Phys.}{A17}{2002}{1003}

\bibitem{lamine:2002}
  \Name{Lamine B., Jaekel M.-T. \and Reynaud S.}
  \REVIEW{Eur. Phys. J.}{D20}{2002}{165}

\bibitem{lamine:thesis}
  \Name{Lamine B.}
  \Book{PhD thesis}
  \Year{2004}, http://tel.ccsd.cnrs.fr/

\bibitem{zeilinger}
  \Name{Hornberger K., Uttenthaler S., Brezger B. \etal}
  \REVIEW{Phys. Rev. Lett.}{90}{2003}{160401};
  \Name{Hackermueller L., Hornberger K., Brezger B. \etal}
  \REVIEW{Nature}{427}{2004}{711}

\bibitem{Hyper00}
  \Book{HYPER, Hyper-precision cold atom interferometry in space}
  \Editor{Assessment study report ESA-SCI}
  \Year{2000}.

\bibitem{gwdetectors}
A list of links is given on the VIRGO website  http://www.virgo.infn.it/facilities/othergw.html

\bibitem{bender:97}
 \Name{Bender P.L. \and Hils D.}
 \REVIEW{Class. Quantum Grav.}{14}{1997}{1439}

\bibitem{grishchuk}
 \Name{Grishchuk L.P.}
 \REVIEW{Sov. Phys. JETP}{40}{1975}{409};
 a recent discussion and references can be found in
 \Name{Grishchuk L.P.} {arXiv:gr-qc/0504018},
 to be published in  \textit{Phys. Uspekhi}.

\bibitem{schutz:99}
 \Name{Schutz B.}
 \REVIEW{Class. Quantum Grav.}{16}{1999}{A131}

\bibitem{maggiore:00}
 \Name{Maggiore M.}
 \REVIEW{Phys. Reports}{331}{2000}{283}

\bibitem{ungarelli:01}
 \Name{Ungarelli C. \and Vecchio A.}
 \REVIEW{Phys. Rev.}{D63}{2001}{064030}

\bibitem{grishchuk:01}
 \Name{Grishchuk L.P., Lipunov V.M., Postnov K.A. \etal}
 \REVIEW{Phys. Uspekhi}{44}{2001}{1}

\bibitem{tourrenc:76}
  \Name{Linet B. \and Tourrenc P.}
  \REVIEW{Can. J. Phys.}{54}{1976}{1129}

\bibitem{landau:book}
  \Name{Landau L. \and Lifchitz E.}
  \Book{Field theory}
  \Editor{Mir}
  \Year{1970}

\bibitem{weinberg:book}
  \Name{Weinberg S.}
  \Book{Gravitation and cosmology}
  \Editor{Wiley}
  \Year{1972}

\bibitem{imry:90}
 \Name{Stern A., Aharonov Y. \and Imry Y.}
 \REVIEW{Phys. Rev.}{A41}{1990}{3436}

\bibitem{borde:01}
  \Name{Bord\'e C.J.}
  \REVIEW{C.-Rendus Acad. Sci. Paris IV}{2}{2001}{509}

\bibitem{chu:01}
  \Name{Peters A., Chung K.Y. \and Chu S.}
  \REVIEW{Metrologia}{38}{2001}{25}

\bibitem{reynaud:2004}
  \Name{Reynaud S., Lamine B., Lambrecht    A. \etal}
  \REVIEW{Gen. Rel. Grav.}{36}{2004}{2271}

\bibitem{chu:94}
  \Name{Weitz M., Young B.C. \and Chu S.}
  \REVIEW{Phys. Rev. Lett.}{73}{1994}{2563}

\bibitem{andersson:2002}
  \Name{Andersson E., Calarco T., Folman R. \etal}
  \REVIEW{Phys. Rev. Lett.}{88}{2002}{100401}

\bibitem{pfau:93}
  \Name{Pfau T., Kurtsiefer C., Adams C.S. \etal}
  \REVIEW{Phys. Rev. Lett.}{71}{1993}{3427}

\bibitem{houde:00}
  \Name{Houde O., Kadio D. \and Pruvost L.}
  \REVIEW{Phys. Rev. Lett.}{85}{2000}{5543}

\bibitem{robert}
  \Name{Boustimi M., Baudon J., Ducloy M. \etal}
  \REVIEW{Eur. Phys. J.}{D17}{2001}{141}

\bibitem{arndt:private}
  \Name{Arndt M.}
  {Private discussion}.



\end{thebibliography}
\end{document}